**On the passive gravitational mass**

A. LOINGER

Dipartimento di Fisica, Università di Milano

Via Celoria, 16 –  20133 Milano, Italy

**Summary.** – A good candidate to the role of passive gravitational mass of a Schwarzschild's incompressible sphere is represented by the *substantielle Masse* of the sphere, which is larger than the active gravitational mass. Presumably this inequality has a general validity, i.e. it is not limited to the homogeneous spheres of incompressible fluids.

PACS 04.20 –  General relativity: fundamental problems and general formalism.

**1.** –  To give in general relativity a proper definition of passive gravitational mass (which, owing to the generalized Galileo's law, coincides with the inertial mass) is not a plain task, see e.g. [1], [2], [3]. According to Møller [1] and Rosen-Cooperstock [3], the passive mass is always equal to the active mass (as in the Newtonian theory), whereas Bonnor [2] finds that the passive mass is always larger than the active one. It seems to me that the second fundamental memoir by Schwarzschild [4] yields a key for solving the puzzle. In this memoir the Author solves the problem of the gravitational  field generated by a  homogeneous sphere of an incompressible  fluid. He defines *two* different concepts of mass of the above sphere, the attractive (or active) gravitational mass and the substantial mass (*substantielle Masse*), which is greater than the active one. While the first concept found a full acceptance in the scientific literature, as is well known, the concept of substantial mass fell into oblivion; only Møller [4] discusses it, but almost *en passant* and  emphasizing  that  for  usual  astronomical  objects,  if  represented  by Schwarzschildian spheres, the substantial mass differs very little from the active mass.

I wish to show that Schwarzschild's substantial mass is a good candidate to the role of passive mass of an incompressible sphere. As remarked by Bonnor [2] (whose definition of passive mass of a Schwarzschild's sphere does not coincide with the substantial mass), it is unclear what are the consequences of the inequality of the active and passive masses for motion of the bodies — inequality which holds likely in general, not only for the mentioned spheres. Presumably these consequences are less drastic than in Newtonian dynamics, by virtue of the conservation laws $T^{jk}{}_{;k} = 0, \ (j,k = 0,1,2,3)$ .

**2.** — According to Schwarzschild [4], the Einstein field of a homogeneous sphere of an incompressible fluid is characterized by the following equations:

*Interior* region:

$$(2.1) \qquad ds^2 = \left(\frac{3\cos\chi_a - \cos\chi}{2}\right)^2 c^2 dt^2 -$$

$$- \frac{3}{\kappa\rho_0}(d\chi^2 + \sin^2\chi \ d\vartheta^2 + \sin^2\chi \ \sin^2\vartheta \ d\varphi^2) \ , \qquad (0 \le \chi \le \chi_a < \pi/2) \ ,$$

where the suffix *a* denotes a value at the surface of the sphere, $\kappa := 8\pi G/c^2$ , and $\rho_0 = T_0^0/c^2$ is the uniform and constant mass density —

*Exterior* region (2*m* coincides with Schwarzschild's $\alpha$ ):

$$(2.2) \qquad ds^2 = \left(1 - \frac{2m}{R}\right) c^2 dt^2 - R^2(d\vartheta^2 + \sin^2\vartheta \ d\varphi^2) - \left(1 - \frac{2m}{R}\right)^{-1} dR^2 \ ,$$

where [5]:

(2.2') 
$$R := (r^3 + \rho)^{1/3} \quad ,$$

(2.2'') 
$$r^3 = (\kappa \rho_0 / 3)^{-3/2} \{(9/4) \cos \chi_a [\chi - (1/2) \sin 2\chi] - (1/2) \sin^3 \chi\} ,$$

(2.2''') 
$$\rho := (\kappa \rho_0 / 3)^{-3/2} \{(3/2) \sin^3 \chi_a - (9/4) \cos \chi_a [\chi_a - (1/2) \sin 2\chi_a]\} \quad ,$$

(2.2$^{iv}$) 
$$2m = (\kappa \rho_0 / 3)^{-1/2} \sin^3 \chi_a = \frac{\kappa \rho_o}{3} R_a^3 \quad .$$

We see from (2.3$^{iv}$) that the active gravitational mass $mc^2/G$ is equal to $(4/3)\pi R_a^3 \rho_0$ .

Evidently the volume of our sphere is

(2.3) 
$$\int_0^{\chi_a} d\chi \int_0^\pi d\vartheta \int_0^{2\pi} d\varphi \ \gamma^{1/2} \quad ,$$

where

(2.3') 
$$\gamma^{1/2} = \left(\frac{3}{\kappa \rho_0}\right)^{3/2} \sin^2 \chi \ \sin \vartheta \ ;$$

thus we have

(2.3'') 
$$V = 2\pi \left(\frac{3}{\kappa \rho_0}\right)^{3/2} (\chi_a - \frac{1}{2}\sin 2\chi_a) \quad ,$$

and the mass $\rho_0 V$ of the sphere (Schwarzschild's *substantielle Masse*) does not coincide with the active mass $(4/3)\pi R_a^3 \rho_0$ ; their ratio is given by

(2.4) 
$$\frac{(4/3)\pi R_a^3 \rho_a}{\rho_a V} = \frac{2}{3} \frac{\sin^3 \chi_a}{\chi_a - (1/2)\sin 2\chi_a} \quad ;$$

In Fig.1 we have a diagram of the function (2.4):

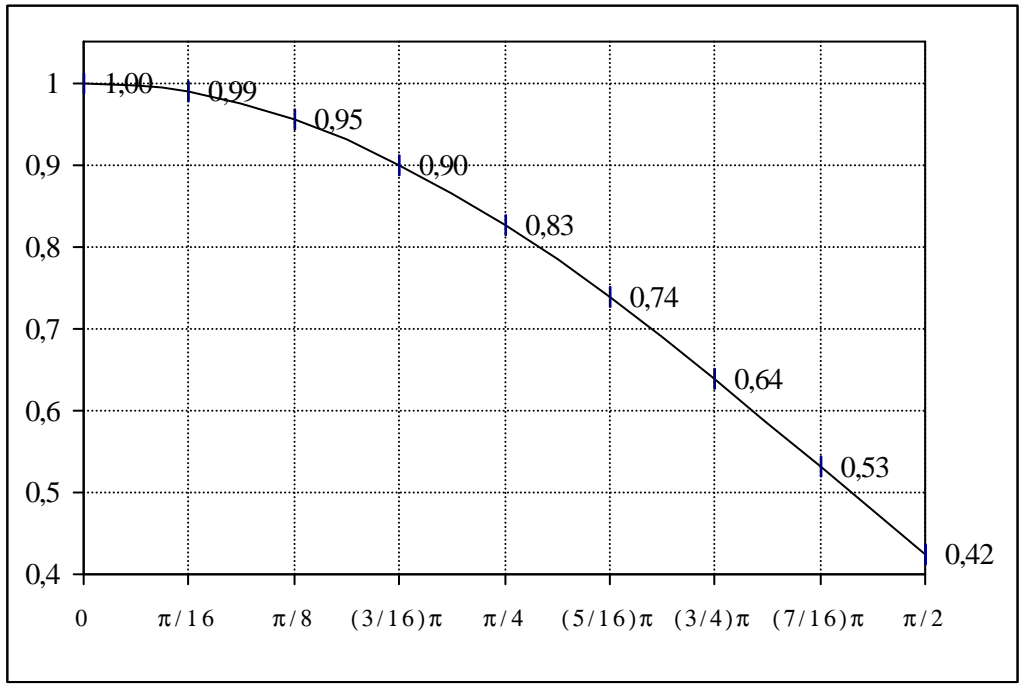

Fig.1 - Diagram of the function $y(x) = \dfrac{2}{3} \dfrac{\sin^3 x}{x - (1/2)\sin 2x}$ , $(0 \le x < \pi/2)$ .

(Note that, in particular, $y[\arccos(1/3)] \approx 0,609$ ).

I propose to consider, *by virtue of its very definition*, the substantial mass $\rho_0 V$ as the passive gravitational mass of the sphere. Bonnor's passive mass [2] is defined by a disputable application of the generalized Galileo's law − as remarked by Rosen and Cooperstock [3]. On the other hand, Møller and Rosen-Cooperstock find that the active and passive masses must coincide, but their proofs rest on a notion of very dubious reliability, the *pseudo* energy-momentum of the gravity field.

**3.** − Externally to *any* matter distribution of spherical symmetry the spacetime interval is given by eq. (2.2), with a constant $\rho$ which depends on the structure of the distribution. It is quite intuitive that *in general* the active mass is different from the substantial mass of the distribution, which we assume as coincident with the passive gravitational mass. The reason of the inequality is "geometrical": the real volume of the distribution is given by a formula (2.3), while the active mass depends on a "Euclidean" volume $(4/3)\pi R_a^3$ .

It is likely that the difference between the active and passive masses is an aesthetic defect of general relativity, whose practical importance should be not great.